\documentstyle[aas2pp4]{article}
\newcommand{\beq}{\begin{equation}}
\newcommand{\eeq}{\end{equation}}

\begin{document}

\title{Interferometric Astrometry of \\ Proxima Centauri and Barnard's Star\\ Using {\it 
Hubble Space Telescope} 
Fine
Guidance Sensor 3: \\
Detection Limits for sub-Stellar Companions \footnote{Based on 
observations made with
the NASA/ESA Hubble Space Telescope, obtained at the Space Telescope
Science Institute, which is operated by the
Association of Universities for Research in Astronomy, Inc., under NASA
contract NAS5-26555} }

\author{ G.\ Fritz Benedict\altaffilmark{2}, Barbara 
McArthur\altaffilmark{2}, D. W. Chappell\altaffilmark{12},
E.\ Nelan\altaffilmark{3}, 
W.\ H.\ Jefferys\altaffilmark{4}, W.~van~Altena\altaffilmark{7}, J. 
Lee\altaffilmark{7}, D. Cornell\altaffilmark{13},
P.~J.~Shelus\altaffilmark{2},
 P.D. Hemenway\altaffilmark{5},	Otto G.\ Franz\altaffilmark{6}, L.\ 
H.\
Wasserman\altaffilmark{6},	R. L.
Duncombe\altaffilmark{11}, D. Story\altaffilmark{9}, A.\ L.\
Whipple\altaffilmark{10}, and L.W.\
Fredrick\altaffilmark{8}}
\altaffiltext{2}{McDonald Observatory, University of Texas, Austin, TX 78712}
 \altaffiltext{3}{Space Telescope Science Institute, 3700 San Martin Dr., Baltimore, MD 21218}
\altaffiltext{4}{Astronomy Dept., University of Texas, Austin, TX 78712}
 \altaffiltext{5}{Oceanography, University of Rhode Island, Kingston, RI 02881}
 \altaffiltext{6}{Lowell Observatory, 1400 West Mars hill Rd., Flagstaff, AZ 86001}
 \altaffiltext{7}{Astronomy Dept., Yale University, PO Box 208101, New Haven, CT 06520}
\altaffiltext{8}{Astronomy Dept., University of Virginia, PO Box 3818, Charlottesville, VA 22903}
\altaffiltext{9}{ Jackson and Tull, Aerospace Engineering Division
7375 Executive Place, Suite 200, Seabrook, Md.  20706}
\altaffiltext{10}{Allied-Signal Aerospace, PO Box 91, Annapolis Junction, MD 20706}
\altaffiltext{11}{Aerospace Engineering, University of Texas, Austin, TX 78712}
\altaffiltext{12}{Environmental Systems Science Center,
University of Reading, PO Box 238, Reading RG6 6AL, UK}
\altaffiltext{13}{Tracor Aerospace, 6500 Tracor Lane, MD 28-4
Austin, TX 78725}



\begin{abstract}
We report on a sub-stellar companion search utilizing interferometric fringe-tracking astrometry acquired with Fine 
Guidance Sensor 3 (FGS 3) on the {\it Hubble Space Telescope}. Our targets were Proxima Centauri and Barnard's Star. We obtain absolute parallax values  for Proxima Cen $\pi_{abs} =0\farcs 7687 \pm 0 \farcs 0003$  and for Barnard's Star $\pi_{abs} =0\farcs 5454 \pm 0\farcs 0003$. 

Once low-amplitude instrumental systematic errors are identified and removed, our companion detection sensitivity is less than or equal to one Jupiter mass for 
periods longer than 60 days for Proxima Cen. Between the astrometry and the \cite{Kur99} radial velocity results we exclude all companions with $M > 0.8M_{Jup}$ for the range of periods $1 < P < 1000$ days. For Barnard's Star our companion detection sensitivity is less than or equal to one Jupiter mass for periods longer than 150 days. Our null results for Barnard's Star are consistent with those of Gatewood (1995). 

\end{abstract}




%

\section{Introduction}
Currently accepted theories predict that planetary systems are a natural
by-product of the formation of stars (\cite{Lis93}). \cite{Bla95} reviews   
the
importance of searches for extrasolar planets to theories of solar system 
formation
and discusses the lack of results to mid-1995. These searches
  have recently succeeded (e.g. \cite{Mar97} and \cite{Coc97}). 
Radial velocity methods have been used for these detections.
However, the derived planetary (or brown dwarf)  masses are lower limits 
because of the unknown orbital inclination. 
These detections include 
planets with minimum masses ranging from half a Jupiter mass 
 (51 Peg, \cite{May95}) 
to more than seven times the mass of Jupiter (70 Vir, \cite{Mar96}). 
All of these objects orbit stars of solar spectral type, although they may tend to 
unusual metal richness (\cite{Gon96}, \cite{Gon98}). Recently \cite{Del98} and ~\cite{Mar98}, 
detected a planetary mass companion to the M4 dwarf star Gl 876. Direct imaging has yielded two  relatively high-mass
objects, the $20 - 50\, M_{\rm  Jupiter}$ brown dwarf, Gl 229B  (Nakajima 
et al.\ 1995) and from Rebolo et al. (1998) G 196-3B ($15 - 40\, M_{\rm  Jupiter}$). Both orbit M stars.

The relative position information provided by astrometry
resolves the mass uncertainty associated with radial velocity detections.
An astrometric search recently evidenced a long-period Jupiter-mass companion to an 
M2 star, Lalande 21185 (\cite{Gat96}).
Past examples
of the successful application of astrometry to the discovery of
stellar-mass unseen companions include Harrington (1977) and Lippincott 
(1977).

Black and Scargle (1982) were the first to point out that Jupiter-like  planets
orbiting M stars might have short periods. They argued that scaling  down 
the pre-planetary nebula thought to accompany the formation of stars  would 
result 
in a
gas-giant forming relatively close to a low-mass star.  Such an object 
would
have a period far shorter than that of Jupiter around  the Sun. This was the 
motivation for our {\it Hubble Space Telescope (HST)} planet search program.
The intervening years have changed this picture to include the formation of
gas-giant planet cores well away from the parent star, even for M stars 
(\cite{Bos95}), with subsequent inward orbital migration (\cite{Lin96}).

We present astrometry of Proxima Cen and Barnard's Star, including 
results of our astrometric searches for brown dwarf and
 planetary-mass companions. Those interested can trace the history of data  acquisition and analysis over the past six years through a series of progress reports (Benedict et al, 1994a, 1994b, 1995, 1997, and 1998c). 

Given the null result of an astrometric companion search carried out by Gatewood (1995) for Barnard's Star, the thirty four observation sets of this field nonetheless have 
significant value. Any systematics 
introduced by {\it HST} and/or FGS 3 should be present in the data for both 
Barnard's Star and Proxima Cen. In this sense Barnard's Star is a control in the 
experiment to detect low-mass companions orbiting Proxima Cen. 

Past direct detection observations include \cite{VBu98}, who observed Barnard's Star at $10\mu$ and established companion upper limits $70 - 80M_{Jup}$ for separations $4 < a < 18$ AU.  A possible companion to Proxima Cen detected by the {\it HST} FOS used in a coronagraphic mode (\cite{Sch98}) is not confirmed by our astrometric data, as was pointed out in that paper. Such a companion must possess an exceptionally low mass to luminosity ratio. A subsequent study using {\it HST} WFPC-2 (\cite{Gol98}) also failed to confirm this companion or find any as bright with separations $0.26 < a < 1.11$ AU ($0\farcs2 < a < 0\farcs85$).

In \S 2 we briefly review the astrometer, FGS 3, and discuss the  data sets. We outline our calibration procedures (\S 3), and present in \S 4 the astrometric modeling of the reference frames. In \S 5  we derive parallaxes and proper motions
for Proxima Cen and
Barnard's Star. In \S 6 we  
explore the astrometric residuals for periodic phenomena indicative of 
companions. In \S 7 we discuss the astrometric results, first the
differences  between
{\it HST} and {\it HIPPARCOS}, and then our  
 mass detection limits that rule out at the 0.1\% level Brown Dwarf companions ($M\le 0.013M_{sun}$) to these two stars. See \cite{OKS99} for a discussion of the mass limits defining Brown Dwarfs.

Tables~\ref{tbl-1} and
\ref{tbl-2} provide aliases and physical parameters for our two science 
targets. The masses are based on the parallax results of this paper (\S 5). From those distances we obtain the V-band absolute magnitudes, M$_{V}$ in Tables~\ref{tbl-1} and \ref{tbl-2}, then obtain masses from the recent lower main sequence mass-luminosity relationship of \cite{Hen99}. Radii are from the models of \cite{Bur93}, confirmed by the CM Draconis results of \cite{Met96}.

We time-tag our data with a modified Julian Date, $mJD = JD - 2444000$, and abbreviate milli-second of arc, mas, throughout.
\section{The Astrometer and the Data} \label{fgs3}
Our goal, 1 mas precision small-field 
astrometry, has been achieved, but not without significant challenges.
Our observations  were
obtained with Fine Guidance Sensor 3 (FGS 3), a two-axis, white-light
interferometer  aboard {\it HST}. 
\cite{Bra91}
provide an overview of the
FGS 3 instrument and \cite{Ben94a} describe the astrometric capabilities 
of FGS 3 and typical data acquisition strategies.

The coverage for both targets suffers from extended gaps, due to {\it HST}
pointing constraints
(described in \cite{Ben93}) and other scheduling difficulties. For Proxima 
Cen the data now include 152 shorter exposures
secured over 4 years (March 1992 to October 1997) and 15 longer 
exposures
(July 1995 to July 1996).  Each
orbit contains from two to four exposures. The longest exposure times
pertain only to Proxima Cen
observations obtained within Continuous Viewing Zone (CVZ) orbits. 
These
specially scheduled orbits
permit $\sim90$ minutes on field, during which  Proxima Cen was not
occulted by the Earth.
See Benedict et al. 1998 (Appendix 1, Table A1) for times of observation and 
exposure times for all 
Proxima Cen astrometry. A total of fifty-nine orbits have 
astrometric 
value. However, data acquired prior to mJD 8988 is of overall lower quality. During this era we had no on-orbit instrumetal astrometric calibration and no independent scale assessment (\S 3.1), no assessment of the time-varying component of our astrometric calibrations (\S 3.2), and employed only rudimentary (constant rate) intra-orbit drift correction (\S 3.3). 

Barnard's Star was monitored for three years (February 1993 to April 
1996),
and observed three times during
each of 35 orbits. Thirty-four of these orbits have astrometric value. 
Exposures range between 24 and 123 seconds duration.
See Benedict et al. 1998 (Appendix 1, Table A2) for times of observation and 
exposure times for all Barnard's Star astrometry.

\section{Data Reduction and Calibration Procedures}
Our data reduction and calibration procedures have evolved since the preliminary 
description given in \cite{Ben94a}. 
To remove systematics from an astrometric reference
frame (see \cite{Ben98a}
for a more detailed discussion), we centroid raw data, removing intra-observation jitter; apply an Optical Field Angle Distortion (OFAD) calibration 
(\cite{Whi95}, \cite{McA97}); apply time dependent corrections to the OFAD (\cite{McA97}); and correct for drift during each observation set (intra-orbit drift). Additionally, we apply a  lateral color correction (depending on star color index) during the orbit-to-orbit astrometric modeling.
\subsection{The OFAD}
Until recently, no  calibration star field with cataloged 1 mas precision astrometry, our desired performance goal, existed. We used FGS 3 to calibrate itself with multiple observations of a distant star field (M35). A distant field was required so that during the 2-day duration of  data acquisition, star positions would not change. We obtained these data in early 1993 and reduced them with overlapping plate techniques to solve for distortion coefficients and star positions simultaneously. An astrometric catalog of 27 stars with 1 mas positions, covering an area
of $1600 \times 700$ arcsec now exists.
 
The aberration of the optical telescope assembly, along with the optics of the FGS,
comprise the OFAD. The largest component of the design distortion, which
consists of several arcseconds, is an effect that mimics a change
in plate scale.
The magnitude of non-linear, low frequency distortions is on
 the order of 0.5 seconds of arc over the FGS field of view.
The OFAD is the most significant source of systematic error in
fringe-tracking astrometry done with the FGS.
We have adopted a pre-launch functional form originally developed by
Perkin-Elmer, the builders of the FGS.
 It can be
 described (and modeled to the level of one millisecond of arc) by the two
dimensional fifth order polynomial:

\begin{eqnarray}
X' = a_{00}  + a_{10}X +a_{01}Y + a_{20}X^{2} 
+a_{02}Y^{2} + a_{11}XY \\
+a_{30}X(X^{2}+Y^{2}) + a_{21}X(X^{2}-Y^{2}) \nonumber  \\
+  a_{12}Y(Y^{2}-X^{2}) + a_{03}Y(Y^{2}+X^{2}) \nonumber \\
+a_{50}X(X^{2}+Y^{2})^{2} + a_{41}Y(Y^{2}+X^{2})^{2}  \nonumber \\
+  a_{32}X(X^{4}-Y^{4}) + a_{23}Y(Y^{4}-X^{4}) \nonumber \\
+a_{14}X(X^{2}-Y^{2})^{2} + a_{05}Y(Y^{2}-X^{2})^{2}  \nonumber \\
\nonumber
\end{eqnarray}
\begin{eqnarray}
Y' = b_{00} + b_{10}X +b_{01}Y + b_{20}X^{2} 
+b_{02}Y^{2} + b_{11}XY  \\
+b_{30}X(X^{2}+Y^{2}) + b_{21}X(X^{2}-Y^{2}) \nonumber \\
+ b_{12}Y((Y^{2}-X^{2}) + b_{03}Y(Y^{2}+X^{2}) \nonumber \\
+b_{50}X(X^{2}+Y^{2})^{2} + b_{41}Y(Y^{2}+X^{2})^{2}  \nonumber \\
+ b_{32}X((X^{4}-Y^{4}) + b_{23}Y(Y^{4}-X^{4}) \nonumber \\
+b_{14}X(X^{2}-Y^{2})^{2} + b_{05}Y(Y^{2}-X^{2})^{2} \nonumber\\
\nonumber
\end{eqnarray}
where X, Y are the observed position
within the FGS field of regard, $X', Y'$
are the corrected position, and the numerical values
of the coefficients $a_{ij}$ and $b_{ij}$ are determined by calibration. The calibration observations required 19 orbits and produced over 570 star position observations. We then employed GaussFit (\cite{Jef88}), 
a least squares
and robust estimation package, to simultaneously estimate the terms in the OFAD equations and the star positions within M35.

\subsection{Maintaining the OFAD}
The FGS 3 graphite-epoxy optical bench was expected to outgas for a period of time after the launch of {\it HST}. This outgassing was predicted to change the relative positions of optical components on the optical bench.  The result of whatever changes were taking place was a change in scale.  The amount of scale change was far too large to be due to true magnification changes in the HST optical assembly. To track these scale-like changes we revisit the M35 calibration field periodically, the ongoing LTSTAB (Long-Term STABility) series. LTSTABs will be required as long as it is desirable to do 1 mas precision astrometry with FGS 3. The result of this series is to model and remove the slowly varying component of the OFAD, so that uncorrected distortions remain below 2 mas for the center of the FGS 3 field of regard. The LTSTAB series is also the diagnostic for deciding whether or not a new OFAD is required.

\subsection{Intra-orbit Drift Corrections}
A major improvement implemented since the earlier report (\cite{Ben94a}) is application of inter-observation (intra-orbit) drift corrections. Over the span of an orbit the positions reported by FGS 1 and FGS 2 for the guide stars change. \cite{Ben98a} shows an observation set with  drift exceeding 30 mas over the span of 36 minutes. The X and Y drift rates are generally dissimilar and usually not constant. The solution to this problem is simple and effective, but imposes additional overhead, reducing the time available within an orbit to observe the  science target. An observation set must contain visits to one or more astrometric reference stars, multiple times during each observation sequence. Presuming no motion intrinsic to these stars over a span of 40 minutes, one determines drift and corrects the reference frame and target star for a  drift generally quadratic with time. As a result we minimize the error budget contribution from drift to typically less than 1 mas.

\subsection{Lateral Color Calibration}
Since this calibration had not previously been applied to our data, we describe it in some detail. Because each FGS contains refractive elements (Bradley et al. 
1991), it is possible that the position measured for a star could
depend on its intrinsic color. Changes in position would depend on star color, 
but the direction of shift is expected to be
constant, relative to the FGS axes. This lateral color shift would be 
unimportant, as long as target and reference stars have
similar color. However, this is certainly not the case for the very red stars, Proxima Cen 
and 
Barnard's Star, hence our interest.
Pre-launch ground testing  (\cite{LAR94}) indicated for FGS 3 a lateral color 
effect predominantly in the X direction, with magnitude approximately 1 mas per unit change in B-V color index. 

An on-orbit test was designed and conducted in December 1991. Due to excessive 
spacecraft jitter (from a combination of the
original solar arrays and insufficient damping in the telescope pointing control 
system) and insufficient knowledge of the
OFAD, the results were inconclusive. We repeated the test in December 1994.  
Analyses  of
these data have yielded a lateral color calibration.
\subsubsection{The Lateral Color Field and Data}
Photometric 
and positional data are presented in Table~\ref{tbl-3} for the four stars in our chosen lateral color test field. We 
obtained POS mode data during three consecutive orbits on mJD 9708 (22 Dec 
1994), placing the four star asterism near each end and in the middle of FGS 3. We observed these additional positions to explore Lateral Color variations within FGS 3. 
At each position HST was rotated $+60\arcdeg$ and $-120\arcdeg$ from nominal roll. 
These large rolls were permitted, since on this date the field (on the ecliptic) 
was at the anti-sun. Of the nine data sets, one was so badly affected by 
spacecraft motion to be unusable. 
\subsubsection{The Lateral Color Model and Results}
Unfortunately, most of the originally proposed reference stars proved too faint and were not observed. The test resulted in enough observations for a single overlapping plate reduction. The eight usable data sets were modeled by 

\beq
X'	=	X + ctx*(B-V) 
\eeq
\beq
	Y'	=	Y + cty*(B-V) 
\eeq
\beq
	\xi = cos(R)*X' - sin(R)*Y' + c  
\eeq
\beq
	\eta = sin(R)*X' + cos(R)*Y + f 
\eeq

solving for R, spacecraft roll; c and f, offsets in X and Y; and global lateral color terms, 
ctx and cty, 
using the overlapping plate techniques described in \cite{Ben94a}. Again using GaussFit 
we find

$ctx = -0.9 \pm 0.2$ and $cty = -0.2 \pm 0.3 $ mas. 

Figure~\ref{fig-1}
presents histograms of the residuals obtained  by applying the model (equations 
1 - 4) to
the lateral color data. To demonstrate that there were no large remaining 
unmodeled  
systematic effects in these data we inspected the X and Y residuals plotted against X and Y 
position within FGS 3.

\section{The Barnard's Star and Proxima Cen Astrometric Reference Frames}  \label{AstRefs}
Figure \ref{fig-2} shows the distribution in FGS 3 pickle coordinates of the thirty-four sets of reference star measurements for the Barnard's Star reference frame. Figure \ref{fig-3} contains the corresponding distribution for the Proxima Cen reference frame. The circular patterns are impressed by the requirement that {\it HST} roll to keep its solar panels fully illuminated throughout the year. 
\subsection{The Model}
From these data we determine the scale and rotation relative to the constraint plate  for each observation set within a single orbit.  Since for Proxima Cen these observation sets span over five years (and over three years for Barnard's Star), we also include the effects of reference star parallax and proper motion. 

Our present model 
\beq
\xi = a*X + b*Y + c -P_x*\pi - \mu_x*t 
\eeq 
\beq 
\eta = d*X +e*Y +f-P_y*\pi - \mu_y*t 
\eeq 

differs slightly from that used previously on the Proxima Cen reference frame (\cite{Ben94a}). X and Y are now corrected for lateral color (equations 3 and 4). The orientation of each data set is obtained from ground-based astrometry. Uncertainties in the field orientations are 0\fdg03 and 0\fdg09 for Proxima Cen and Barnard's Star. We obtain the parallax factors, $P_x$ and $P_y$ from a JPL Earth orbit predictor (DE200, \cite{Sta90}). 
Finally we constrain $\Sigma \mu = 0$ and   
$\Sigma \pi = 0$ for the entire reference frame.

Below we discuss the results of this modeling for the Proxima Cen and Barnard's Star reference frames in parallel.
\subsection{Scale Stability}

As in Benedict (1994) we form a scale-like parameter, S, where
\beq
S = (ae-bd)^{1/2}
\eeq
and $a, b, d, e$  are the coefficients determined in equation 5 and 6.
S for both the Barnard's Star and Proxima Cen reference frames is plotted against time in days in Figure \ref{fig-4}. Error bars are the internal error, $\sigma_{S}$, derived from $\sigma_{a,b,d,e}$. Since both sets have an arbitrary scale zero point, $S_o$, a shift has been applied to bring agreement in the mean scale in the range $9000 \le mJD \le 10300$. 

The most obvious feature is coincident with our early 1993 OFAD determination, at which time we moved from a predicted (from ground-based measurements at the manufacturing facility) to a measured OFAD. Servicing missions (SM1 and SM2) have the potential to disrupt the position of FGS 3 within HST and introduce new volatiles. Neither SM1 nor SM2 seems to have an identifiable effect in Figure \ref{fig-4}. 

From the scatter proximate to each observation date, it is obvious that the scale is indeterminate at the $5\times10^{-5}$ level. 
Each observation set is obtained with the primary science targets (Proxima Cen and Barnard's Star) positioned within 5 arcsec of the pickle center. Thus, the effect of this scale indeterminancy on their calibrated positions is on order 0.3 mas.

Since a search for periodic astrometric phenomena is a primary goal of this investigation, the quasi-periodic behavior of $S$ should be characterized with the same tools we will use on Proxima Cen and Barnard's Star. Our periodogram tools are described in \S 6 and Appendix A1.1. Figure \ref{fig-5} contains a periodogram characterizing the temporal behavior of $S$. Note the large amount of low-frequency power for $800<f< 2000 d^{-1}$. If some uncorrected fraction of this small variation is detectable in the measurements of our primary science targets, we would expect to see it only in Proxima Cen. Observations of Barnard's Star are all post- mJD 8988, the location of the major break in the scale trend, pre- and post-OFAD. 

\subsection{Reference Frame Results}
Tables \ref{tbl-4} and \ref{tbl-5} provide identifications, magnitudes, colors and relative positions for all reference stars. The relative positions within each frame have been rotated to J2000 RA and Dec, at epoch 1996.42 for the Barnard's Star field, 1996.79 for the Proxima Cen field.
 The Proxima Cen reference frame star magnitudes and colors are from \cite{Ben94a}, as is the position of the star relative to which all positions are referenced, star ID-15. The Barnard's Star reference star magnitudes and colors were obtained using the 0.8m CCD Prime Focus camera (~\cite{Cla92}) at McDonald Observatory. Positions are referenced to star ID-30, whose position is from the Guide Star Selection System Catalog (\cite{Las90}). 

Having applied the model (equation 5 and 6) we form histograms of the residuals
(Barnard's Star, Figure~\ref{fig-6}; Proxima Cen, Figure~\ref{fig-7}). No observations have been removed as outliers, but the overlap was carried out using the robust estimation technique described in Jefferys (1988). From these histograms we conclude that our per-observation precision is 0.8 mas in X and 1.1 mas in Y.

\subsection{Assessing Reference Frame Residuals}
The OFAD reduces as-built {\it HST} telescope and FGS 3 distortions with magnitude $\sim1\arcsec$ to below 2 mas (\cite{McA97}) over much of the FGS 3 field of regard. From the histograms (Figures~\ref{fig-6} and~\ref{fig-7},~above) we have obtained correction at the $\sim 1$ mas level in the
region available at all {\it HST} rolls (an inscribed circle centered on the pickle-shaped FGS field of regard). To determine if there might be unmodeled, but eventually correctable, systematic effects at the 1 mas level, we plotted the Barnard's Star and Proxima Cen reference frame X and Y residuals against a number of spacecraft, instrumental, and astronomical parameters. These included X, Y position within the pickle; radial distance from the pickle center; reference star V magnitude and B-V color; and epoch of observation.  We saw no obvious trends, other than an expected increase in positional uncertainty with reference star magnitude.

\section{Modeling the Parallax and Proper Motion of Barnard's Star and Proxima Cen}
Once we have determined plate constants from applying equations 7 and 8 to the reference frames, we apply the transformations to the Proxima Cen and Barnard's Star measurements and solve for relative parallax and proper motion. At this step the lateral color correction is differential. The correction is based on the color difference between the target and the average color of the reference frame. Our relative parallax and proper motion results are presented in Table~\ref{tbl-7}. 

Every small-field astrometric technique requires the following step; a correction from relative to absolute parallax. This correction is required because the reference frame stars have an intrinsic parallax. Ideally, all reference stars would be more distant than the desired science target star parallax precision, a situation rarely provided by nature. To obtain science target parallaxes precise at the 0.5 mas or better level requires this correction. Faint (hence, in general, distant) reference stars have smaller corrections. CCD techniques (e.g., ~\cite{Mon92}) provide an extremely faint reference frame ($V\sim18$), but a relatively small field of view. FGS 3 provides a large field of view and access to moderately faint reference stars ($V>16$). 

We adopt the corrections discussed and presented in the Yale Parallax Catalog (\cite{WvA95}, Section 3.2, Fig. 2, hereafter YPC95). Entering YPC95, Fig. 2, with the Barnard's Star galactic latitude, $l = 13\fdg99$ and average magnitude for the reference frame, $<V_{ref}>=14.2$, we obtain a correction to absolute of 0\farcs0010.

As a test of the validity of this correction we (Cornell and Chappell) have obtained B, V photometry of the Barnard's Star reference frame stars using the 0.8m CCD Prime Focus camera (\cite{Cla92}) at McDonald Observatory. Classification spectra were obtained by one of us (Lee) with the WIYN Observatory\footnote{The WIYN Observatory is a joint facility of the University of
Wisconsin-Madison, Indiana University, Yale University, and the National
Optical Astronomy Observatories.}
~multiobject spectrograph (MOS/Hydra). These data were reduced and
then classified (spectral type and luminosity class) on the MK system by Lee. 
We find that the colors, magnitudes, and spectral types yield a reference frame whose bulk distance properties agree with the YPC95 prediction. Table ~\ref{tbl-6} collects reference star identification numbers; V magnitudes; spectral types and luminosity classes; \bv color indices; absolute magnitudes, M$_V$, and intrinsic color indices, (B-V)$_o$  from the color {\it vs} spectral type and luminosity class listed in  Lang (1992); and resulting color excesses, {E$_{B-V}$. Adopting
$R = 3.1= A_V / E_{\bv}$ 
we obtain the listed total V-band absorptions, $A_V$ and, finally,  the distance moduli, m-M, distances in pc, and parallaxes listed in the last two columns. The average parallax of the reference frame is $<\pi>$ = 1.2 mas.
The difference, 0.2 mas, between this and the YPC95 value, 1.0 mas, provides an estimate of the error in the correction to absolute.

For Proxima Cen ($l = -1\fdg93$ and $<V_{ref}>=14.5$) we obtain a correction,
+1.0 mas .

Applying these corrections results in the absolute parallaxes listed in Table~\ref{tbl-7}. The final formal uncertainties include the estimated error in the correction to absolute, 0.2 mas,  RSS-ed with the relative parallax errors.
We also list absolute parallax and proper motion from the {\it HIPPARCOS} catalog (\cite{Per97}) and the weighted, average parallax from YPC95.

\subsection{The Secular Acceleration of Barnard's Star}
	With a large proper motion and parallax ($\mu = 10\farcs368 y^{-1}$, $\pi = 0\farcs5454$) and large negative radial velocity ($RV = -106.8 km s^{-1}$), Barnard's Star is expected to evidence a secular acceleration in the declination component of the proper motion.
This perspective effect is characterized by
\beq
\dot \mu  = -2.05 \mu\pi RV \times 10^{-6}  
\eeq
and yields for Barnard's Star $\dot \mu = 0\farcs0012 yr^{-2}$.
Gatewood (1995) detected and confirmed the magnitude of this effect with 6 years of data. To determine if our data are sensitive to this effect we modify equation 8 thusly,
\beq 
\eta = d*X +e*Y +f-P_y*\pi - \mu_y*t - \dot \mu_y * t * t
\eeq 
and find $\dot \mu_y = 0\farcs0012 \pm 0\farcs0004 yr^{-2}$, a $3-\sigma$ detection from
3.3 y of data. 

Proxima Cen, with $\mu = 3\farcs851 yr^{-1}, \pi
= 0\farcs7687$, and 
$RV = -21 km s^{-1}$, has a predicted $\dot \mu =  0\farcs00013 yr^{-2}$, undetectable by {\it HST}.

\section{Astrometric Companion Detection}
Applying the proper motion and parallax model (equations 7 and 11 for Barnard's Star, equations 7 and 8 for Proxima Cen) yields residuals for Barnard's Star and Proxima Cen. We wish to test for the existence of periodic behavior in these residuals, astrometric perturbations indicative of planetary-mass companions.

\subsection{Forming Normal Points and Associated Uncertainties}
Every Barnard's Star orbit contained at least three observations. Before subjecting the residuals to a period-finding procedure we average the multiple observations obtained at each epoch (the science target observations collected during a single orbit) to form normal points. This step removes the high-frequency power for $1/f < 0.014 d$, the typical time interval between intra-orbit observations.  Each normal point has an associated uncertainty that is the standard deviation of the intra-orbit observations. Resulting normal points and uncertainties are shown in Figure ~\ref{fig-8}.

For Proxima Cen only those orbits acquired for $t > mJD ~9924$ contained multiple observations of the science target. However, prior to that epoch the frequency of observation was far greater. Consequently, we trade dense temporal coverage at earlier epochs for more realistic normal points and uncertainties by averaging over 3-10 days. This approach results in the 28 normal points plotted in Figure~\ref{fig-9}. 

To summarize,  Barnard's Star and Proxima Cen (for $t > mJD ~9924$) normal points are always formed from data acquired within one orbit. Proxima Cen normal points prior to $ mJD ~9924$ are formed from data acquired over multiple orbits. 

\subsection{Searching for a Perturbation}

The most obvious perturbation search is visual inspection of the residuals plotted against time. We look for trends or obvious periodicities. The residuals for Barnard's Star (Figure~\ref{fig-8}) appear random. The residuals for Proxima Cen (Figure~\ref{fig-9}) do not. We do not have a seamless set of observations for Proxima Cen. Data prior to mJD 8988 had no OFAD, no OFAD maintenance (LTSTABS), and only rudimentary (constant rate) drift correction. If we remove those data prior to mJD 8988 and redo the entire analysis, the residual vs. time plot becomes far flatter. Additionally a weighted fit of a parabola to all the data in 
Figure~\ref{fig-9} yields statistically insignificant curvature.

We next search for more subtle variations. Since the HST observations of 
Proxima Centauri and Barnard's Star are unevenly distributed in time,
traditional Fourier transform techniques which assume regularly spaced
data are inappropriate for period searching.  The usual tactic of clamping the data to zero during large gaps in the
sampling often leads to excessive power on scales of the lengths of the gaps
(Press et al.  1992).
Also, if we interpolated the data onto a regularly spaced grid,
information would be lost and the sought signal effectively degraded. 

Schuster (1905) introduced the periodogram as a simple way of
handling data sets with non-uniform sampling such that each data point 
receives
equal weight. This approach, described in Appendix A1.1, was used in a successful search for photometric variability of Proxima Cen and Barnard's Star (\cite{Ben98b}).

In Figures ~\ref{fig-10} and~~\ref{fig-11}, normalized periodograms are presented for the positional data (the normal points in Figures ~\ref{fig-8} and~~\ref{fig-9}) for Barnard's Star and 
Proxima Cen. Superimposed on each spectrum are horizontal lines of constant
false-positive rates computed using Equation A-4. Barnard's Star evidences  large power at a frequency equivalent to one-half year and Proxima Cen has a significant signature at the lowest frequencies examined for perturbations.

Before further exploring the causes for the strong signatures in the periodograms, we wish to see if these signatures persist for an alternate spectral analysis technique. A Bayesian spectrum analysis technique developed by Bretthorst (1988) and Jaynes (1987) uses probability 
theory to approach the problem of 
signal detection (see Appendix A1.2). This approach is most beneficial where the signal-to-noise ratio
is small. Figure ~\ref{fig-12} contains the resulting Bayesian spectral
estimate for Barnard's Star. In this case the false-positive levels are estimated using the Monte-Carlo method discussed in Appendix A1.2. Noting that the Bayesian and normalized periodogram approaches produce similar results, we will use normalized periodograms in the rest of this paper.

\subsection{Significant Peaks in the Power Spectra}
Most of the power in the Barnard's Star periodogram occurs at $f\sim0.5 y^{-1}$. The Barnard's Star field is unique in one respect, motion across the pickle.
It is our operational constraint to place the science target always within a few arcsec of the pickle center. Barnard's Star has a proper motion in excess of $10\arcsec yr^{-1}$. This combination dragged the reference frame over 30\arcsec ~across the pickle during our monitoring. Consequently, the reference star positions were shifted by 30\arcsec across the pickle, while the Sun constraint rotated the field several times through $360\arcdeg$. Thus, the plate constants derived will likely have periodic OFAD deficiencies buried in them, causing the strong signal in the periodogram.

Both parallax and spacecraft roll correlate with an integral fraction of one year. Figure \ref{fig-13} shows x and y normal points plotted against spacecraft roll (roll $\bmod 180\arcdeg$). There is a clear systematic effect in Y, linear with roll. In X the effect is quadratic with roll. If we substitute the x and y residuals (CorrNP) to these linear fits as our new normal points, the resulting power spectra (Figure~\ref{fig-14}) no longer contain significant power at $f\sim0.5 y^{-1}$. This empirical correction renders our search insensitive to perturbations with 1/2 year 
periods ($f \sim 0.006 day^{-1}$).

With a proper motion roughly one-third that of Barnard's Star, the Proxima Cen periodogram might also be expected to have some signal at $f \sim 0.006 day^{-1}$. However, it does not. The most significant peak in the Proxima Cen periodogram (Figure ~\ref{fig-11}) occurs at the lowest frequency we are able to probe, $f \sim 2000 d^{-1}$. The periodogram is detecting the secular drift in the residuals seen in Figure~\ref{fig-9}.
Comparing the periodograms for scale (Figure~\ref{fig-5}) and the astrometry residuals (Figure~\ref{fig-11}), the peak in the astrometric residuals periodogram has a frequency nearly identical to the peak in the scale variation
periodogram. The amplitude of the secular trend periodogram is much less, indicating possible incomplete correction of scale-like variations.

\subsection{Detection Limits from Power Spectra}

Monte-Carlo simulations were used to determine angular perturbation limits for companions that could hide in these power spectra. For each signal period and amplitude, 2000 simulated data sets of a sinusoidal signal with additive uncorrelated Gaussian noise were generated. The noise variance and sampling pattern were chosen to match the reference star observations (\S4.3). The probability that a spectral peak with amplitude $S_A$ could have arisen from noise is given by false positive rate $P_{fp}(S_A)$. We define a miss rate, $P_{miss}(S_A)$ as the fraction of trials
for which the largest spectral peak either falls below a given spectral amplitude, $S_A$, or occurs at another frequency due to noise fluctuations. The derived perturbation amplitudes are those in which 95\% of the simulations produced power in excess of that corresponding to a 1\% false-positive rate. The detection amplitudes represent a limit on the smallest underlying signal which could give rise to a given spectral amplitude. We find that the detection amplitudes are effectively insensitive to perturbation frequency. For Barnard's Star we obtain 1.25 mas, and for Proxima Cen, 1.0 mas for perturbation frequencies $f < 1500 d^{-1}$.

\section{Discussion of Astrometric Results}
\subsection{The Parallax and Proper Motions of Proxima Cen and Barnard's Star}
As shown in Table~\ref{tbl-7}, we find parallaxes for Proxima Cen and Barnard's Star that differ substantially from values found by {\it HIPPARCOS}. The YPC95 parallax values are weighted averages of four and six independent determinations for Proxima Cen and Barnard's Star, respectively. All values and associated errors are plotted in Figure~\ref{fig-15}, showing good agreement between the {\it HST} and YPC95 parallaxes. The 3.6 and 3.9 mas differences between {\it HST} and {\it HIPPARCOS} parallaxes may be explained in part by {\it HIPPARCOS} zonal errors (e.g.,~\cite{Nar99}). 

The proper motion differences can, in part, be attributed to the {\it HST} values
being relative to a local, not a global, reference frame. That these proper motion differences are at the 1 mas level, demonstrates agreement between the {\it HST} and
{\it HIPPARCOS} scales to within one part in 4000. Scale differences cannot explain the parallax differences.
\subsection{Companion Mass Detection Limits}
The only confirmed planetary mass companion found thus far for an M star has a  short period (Gl 876, $P\sim61^d$, \cite{Del98} and \cite{Mar98}). This may be partly a selection effect, since M stars have been studied extensively with high-precision radial velocity techniques for only  a few years. Nonetheless it is interesting that Figures~\ref{fig-14} and~\ref{fig-11} show no significant power for frequencies $60 < f <12 days^{-1}$. Neither is there power at the stellar rotation rates ($Prot_{Barn}\sim130^d$, $Prot_{Prox}=83.5 \pm 0.5^d$) inferred from the FGS photometry described in \cite{Ben98b}. This is an expected result, since the star spots, with 1-2\% contrast would not be expected to perturb the observed photocenters of Proxima Cen and Barnard's Star.

Our detection limits are expressed in angular measurements.   We require  primary star masses and distances to translate to mass detection limits. We estimate the masses of Proxima Cen and Barnard's Star from the mass-luminosity relationship of Henry et al. (1999). The masses are averages of a mass estimate from M$_V$ derived from our parallax values (Table~\ref{tbl-5}  and V,~\cite{Leg92}) and a B-V =f(M$_V$) relationship (\cite{Hen99}), where B-V is also from Leggett (1992).
We adopt $M_{Prox}= 0.11M_{\sun}$ and $M_{Barn}= 0.16M_{\sun}$. We wish to estimate the mass of a planetary companion, $M_{p}$, from detectable values of $\alpha'$, the semi-major axis of  perturbation orbit in arcsec. From 
\beq
(M_{*}+M_{p})P^2 = a^3
\eeq  
\beq
\alpha M_{*} = a M_{p}
\eeq
\beq
\alpha'(\arcsec) = \alpha(AU) \pi
\eeq
 with P in years, a and $\alpha$ in AU, and M in solar masses, we derive $ M_{p}$ able to produce the angular perturbations detectable at a 1\% false-positive level (Barnard's Star, 1.25 mas; Proxima Cen, 1.0 mas). Our results for Barnard's Star and Proxima Cen are shown in
Figures~\ref{fig-16} and ~\ref{fig-17}. On the latter we show detection limits from the Proxima Cen radial velocity companion search program of K\"{u}rster et al. (1999), showing the complementarity of the two searches.

While fairly extended in period coverage, our results are less so when stated in orbit size, as shown on the top axes of Figures~\ref{fig-16} and ~\ref{fig-17}. How far from the primaries our techniques probe bears directly on comparison with searches carried out by more direct techniques, particularly camera observation of the neighborhoods of these stars. A Saturn-mass companion with $P=2000^{d}$ would orbit Proxima Cen at $a\sim1.5$ AU ($\sim1\farcs2$). A half-Jupiter mass companion orbiting Barnard's Star with $P=1000^{d}$, would have $a\sim1.1$ AU ($\sim0\farcs58$). There is satisfactory overlap for Proxima Cen with the camera observations of ~\cite{Gol98}, ~\cite{Lei97}, and ~\cite{Sch98}. For Barnard's Star there remains a gap, since~\cite{VBu98} reach only inward to 4 AU.

\section{Conclusions}

1. ~ FGS 3, a white light interferometer on {\it HST} used in a fringe-tracking (POS) mode, produces 1 mas precision astrometry.

2. Attaining this precision requires multiple revisits to reference stars during an orbit, necessary to remove short time-scale temporal positional drift.

3. Longer time-scale instrumental changes are monitored and removed through periodic revisits to our primary astrometric calibration field, M35.

4. These techniques yield parallaxes for Barnard's Star and Proxima Cen  with precision better than 0.4 mas. {\it HST} parallaxes differ from the {\it HIPPARCOS} determinations by 3.9 and 3.6 mas, respectively. The {\it HST} determination agrees with the YPC95 value within the errors for Barnard's Star. The secular acceleration of Barnard's Star was determined
at the 3-$\sigma$ level in 3.3 years of observation and agrees with past determinations.

5. We have examined the Barnard's Star and Proxima Cen astrometric residuals for periodic perturbations due to planetary mass companions, using  normalized periodograms and Bayesian spectral analysis. 

6. We find statistically significant power at one-half year for Barnard's Star. The Barnard's Star residuals correlate with spacecraft roll. We tentatively identify this systematic effect with reference frame motion of over 30\arcsec ~across the pickle during our monitoring. Our empirical correction for this sytematic error renders us insensitive to companions with periods $P\sim 0.5^y$. The  periodogram of the corrected data contains no significant power at any inspected frequencies.

7. For Proxima Cen we find statistically significant power at $f\sim 2000 d^{-1}$. 
The source of this power can be seen in the residuals plotted against time. Much of the non-linearity comes from lower quality data secured early in our program. We point out the similarity of this signature with that found in a periodogram of scale-like variations in FGS 3. Together they support the probable identification of a systematic error, not a long-period, low-mass companion.

8. After discarding the signals from systematic errors,  Monte Carlo simulations indicate that a 1 mas amplitude perturbation with a frequency range $2000 d^{-1} > f >30 d^{-1}$ would be found in the Proxima Cen residuals at a 1\% false positive level, with a 5\% miss rate. Corresponding sensitivity for Barnard's Star is 1.25 mas.

9. For Proxima Cen, assuming $M_{Prox} = 0.11M_{\sun}$, we detect no companions more massive than Jupiter with orbital periods $50^{d} < P < 1000^{d}$. For $P>400^{d}$ our detection limit is less than the mass of Saturn. These null results confirm the direct observations of \cite{Lei97} and \cite{Gol98}. They weaken the interpretation of the FOS coronagraphic observations of \cite{Sch98}. Between the astrometry and the \cite{Kur99} radial velocity results we exclude all companions with $M > 0.8M_{Jup}$ for the range of periods $1 < P < 1000$ days.

10. For Barnard's Star, assuming $M_{Barn} = 0.16M_{\sun}$, we detect no companions more massive than twice Jupiter with orbital periods $50^{d} < P < 150^{d}$. For $P>150^{d}$ our detection limit is less than the mass of Jupiter.
\acknowledgments

We thank an anonymous referee for comments and suggestions that greatly improved the form and content of this paper. We thank Artie Hatzes and Bill Cochran for
discussions and draft paper reviews and Melody Brayton for paper
preparation assistance. Denise Taylor provided crucial scheduling
assistance at the Space Telescope Science Institute. Jeff Achterman and Howard Coleman provided support (scientific and morale) during the earliest phases of this project. This research has made use of NASA's Astrophysics Data
System Abstract Service and the SIMBAD Stellar Database inquiry and retrieval system. Support for this work was provided by NASA through grant 
GTO NAG5-1603 from NASA Goddard Spaceflight Center, and grants GO-06036.01-94A and GO-06768.01-95A from the Space Telescope 
Science Institute, which is operated
by the Association of Universities for Research in Astronomy, Inc., under
NASA contract NAS5-26555.
\clearpage

  \renewcommand{\theequation}{A-\arabic{equation}}
  \setcounter{equation}{0}  
  \section*{APPENDIX- Perturbation Search Techniques}  
The standard periodogram
when applied to non-uniformly sampled data has some
unfortunate statistical shortcomings, including
a sampling-dependent probability distribution for pure noise samples
and non-translational invariance.  Both of these 
problems
are cured by a modified version of the periodogram introduced by
Scargle (1982).  We used this modified periodogram to analyze the HST 
data
and briefly review the method in the next section.  We also used a 
technique called Bayesian spectrum analysis which is based on probability 
theory (see Bretthorst 1988).  This method provides a way of estimating
the periodogram defined by Equation 8, but is more robust for very noisy
data sets.  

\subsection*{Appendix A1.1 - The Normalized Periodogram}
Scargle (1982) introduced a new normalization for the periodogram which 
produces an exponential probability distribution for random Gaussian noise
independent of the sampling pattern.
He also included a time delay, defined below, which solves the translational
invariance problem mentioned above.
With these modifications, the ``normalized'' periodogram is:
\beq 
P_{N}(\omega) \equiv 
\eeq
\begin{eqnarray} \frac{1}{2} \left\{ \frac{[\sum_{j} y_j \cos \omega 
(t_j - \tau)]^2}{\sum_j \cos^2 \omega (t_j - \tau)} + \frac{[\sum_{j} y_j 
\sin \omega (t_j - \tau)]^2}{\sum_j \sin^2 \omega (t_j - \tau)}\right\} \nonumber
\end {eqnarray}
where the delay, $\tau$, is defined by 
\beq
\tan (2 \omega \tau ) = \frac{\sum_j \sin 2 \omega t_j}{\sum_j \cos 2 
\omega t_j}.
\eeq
With the inclusion of the time delay, the normalized periodogram becomes 
equivalent to least-squares fitting of the data to sine waves.  

The resulting exponential form of the probability distribution 
is convenient since the false alarm probability takes a 
particularly simple form.  For uncorrelated Gaussian noise, the 
probability, $P_{\rm fp}$, that no peaks in the power
spectrum have amplitudes greater than, $S_A$, is 
\beq
P_{\rm fp} = 1 - (1 - {\rm e}^{-S_A})^{N_{\rm eff}}
\eeq
 where $N_{\rm eff}$ is the number of
independent frequencies searched (Scargle 1982).

\subsection*{Appendix A1.2 - A Bayesian Approach}

A Bayesian spectrum analysis technique developed by Bretthorst (1988)
and Jaynes (1987) starts with the hypothesis $H$ that 
a periodic signal exists in the a data set $D$.  
The goal of signal detection is to find the 
probability $P(H|D,I)$
that the hypothesis is correct given the data and any prior information $I$
that might exist, e.g.\ noise statistics.  Bayes' theorem relates this
probability to other calculable probabilities as follows:
\beq
P(H|D,I) = \frac{P(H|I) P(D|H,I)}{P(D|I)},
\eeq
where $P(H|I)$ is the prior probability of the hypothesis, e.g. if one
hypothesis is more likely than another based on, say, theoretical grounds;
$P(D|I)$ is the prior probability of the data which for this analysis is
just a normalization constant; and $P(D|H,I)$ is the direct probability of 
the
data (also called the likelihood function) and is the probability that the
data set could be produced given the hypothesis and the prior information.

The hypothesis for the planet search is that a sinusoidal signal
defined by $f(t) = B_1 \sin \omega t + B_2 \cos \omega t$ exists in the
data.  The probability that this 
hypothesis is true may be written as $P(\omega,B_1,B_2|D,I) \equiv 
P(H|D,I)$.
However, since only a detection is required, we really only want the
probability that a signal with frequency $\omega$ exists regardless of
the amplitude and phase, i.e. $P(\omega|D,I)$.  BJ refer to the
unwanted parameters $B_1$ and $B_2$ as ``nuisance'' parameters and 
demonstrate that
they can be eliminated from the calculation by a simple integration:
\beq
P(\omega|D,I) = \int \int P(\omega,B_1,B_2|D,I) {\rm d}B_1{\rm d}B_2
\eeq
where $P(\omega,B_1,B_2|D,I) \equiv P(H|D,I)$ is given by Equation 12.  
They assume that $B_1$ and
$B_2$ have flat probability distributions and include this assumption in
the prior $I$.  This method of reducing the dimensionality of the problem
is of course very useful useful for models which
contain a large number of parameters.

Bretthorst and Jaynes  show that since the difference between the model and the data is
ideally just the noise, the likelihood function depends on the sum of
the squares of the difference between the data and the model function 
$f(t)$.
By substituting the model $f(t)$ into the resulting likelihood function,
expanding appropriate terms, and 
integrating over the nuisance parameters, they show that the desired 
probability is related to the periodogram defined in Equation A-1 by:
\beq
P(\omega|D,I) \propto \left[1 - \frac{2 C(\omega)}{N 
\bar{y}^2}\right]^{(2-N)/2}
\eeq
where $\bar{y}$ is the mean of the data.  In the derivation of
this expression it is assumed that only the shape of the noise probability
distribution is known and not noise variance.
One of the advantages of this technique is that it provides an internal
estimate of the accuracy of its predictions; a feature which direct
periodogram and Fourier transform methods lack. 

Since we are not aware of a corresponding analytic form for false positive rates for the Bayesian
technique, Monte-Carlo simulations were employed to obtain those plotted in
Figures ~\ref{fig-12}. Numerical simulations were first used to look for frequency-dependent 
false positive variations of the Bayesian spectral estimate.  This was done
to insure that the non-uniform spacing of the data points could not interact
with the noise to produce excessive power at any of the frequencies of
interest.  Groups of artificial data sets were created using the sampling
times of the observations and contained only uncorrelated Gaussian noise.  
The
variance of the noise was chosen to be the estimated noise variance of
each of the observations.  Bayesian spectra were computed for each
artificial data set over the range of frequencies of interest.  Then,
the frequency and amplitude of the largest peak was recorded for each 
spectra.
In order to estimate the distribution of the largest peaks as a function
of frequency, the frequencies of the peaks were collected into $N_{\rm 
eff}$
bins where $N_{\rm eff}$ is the number of independent frequencies in the
spectral region of interest.

Figure ~\ref{fig-12} shows the results of this analysis for the sampling times of 
the Proxima Centauri positional observations.  2000 artificial data sets
and spectra were created and 40 bins were used.
The bottom panel of the figure shows the spectral amplitude corresponding 
to
false positive rates of $0.1\%$, $0.3\%$, and $1.0\%$ over frequencies
of $40^{-1}$ to $400^{-1} {\rm days}^{-1}$.
In the top panel the relative number of peaks are shown over the same
frequency range.  Since the variations in this quantity are consistent
with counting statistics, we conclude that the false positive rates are
essentially frequency independent over range of frequencies of interest.
Thus, the false positive rates for the Bayesian technique shown in
Figures ~\ref{fig-12} were based on the
probability distribution of peak amplitudes found over the entire 
spectral range of interest.


\clearpage
\begin{center}
\begin{deluxetable}{lll}
\tablecaption{Proxima Cen = $\alpha$ Cen C = GJ 551 = V645 Cen = HIP 70890
 \label{tbl-1}}
\tablewidth{0in}
\tablehead{\colhead{Parameter} &  \colhead{Value}&
\colhead{Reference}}
\startdata
V & 11.09$\pm$0.03 & \cite{Leg92} \nl
\bv & 1.90 $\pm$ 0.04 & \cite{Leg92} \nl
$M_{V}$ & $15.60\pm0.1$ & from  $\pi$, this paper \nl
Sp.T. & M5Ve &    \cite{Gli91}\nl
$M_{\rm Prox}$ & $0.11M_{\sun} \pm 0.01 $  &\cite{Hen99}\nl
$L_{\rm Prox}$ & $0.001L_{\sun}$  &\cite{Lie87}\nl
$R_{\rm Prox}$ &  $0.14R_{\sun}$& \cite{Bur93}
\enddata
\end{deluxetable}
\end{center}

\begin{center}
\begin{deluxetable}{lll}
\tablewidth{0in}
\tablecaption{Barnard's Star = GJ 699 = G140-24 = LHS 57 = HIP 87937œ
 \label{tbl-2}}
\tablehead{\colhead{Parameter} &  \colhead{Value}&
\colhead{Reference}}
\startdata
V & 9.55$\pm$0.03 & \cite{Leg92} \nl
\bv & 1.73 $\pm$ 0.04 & \cite{Leg92} \nl
$M_{V}$ & $13.3\pm0.1$&from $\pi$, this paper \nl
Sp.T. & M4Ve &   \cite{Kir94}\\
$M_{\rm Barn}$ & $0.16M_{\sun}\pm 0.01 $  &\cite{Hen99}\nl
$L_{\rm Barn}$ & $0.0046L_{\sun}$  &\cite{Hen93}\nl
$R_{\rm Barn}$ &  $0.19R_{\sun}$&\cite{Bur93} \nl
\enddata
\end{deluxetable}
\end{center}

\begin{center}
\begin{deluxetable}{ccccc}
\tablewidth{0in}
\tablecaption{Lateral Color Calibration Stars \label{tbl-3}}
\tablehead{\colhead{ID} &  \colhead{V}&
\colhead{\bv}&\colhead{RA (\arcdeg)}&\colhead{Dec (\arcdeg)}}
\startdata
1&10.55&1.92&89.8712&22.5804\nl
2&10.42&0.18&89.8640&22.5777\nl
6&14.25&0.81&89.8861&22.5775\nl
9&14.03&1.51&89.8734&22.5629\nl
\enddata
\end{deluxetable}
\end{center}

\begin{center}
\begin{deluxetable}{cccccccc}
\tablewidth{0in}
\tablecaption{Barnard's Star Reference Frame  \label{tbl-4}}
\tablehead{\colhead{Star} &  \colhead{ID}&
\colhead{V}&   \colhead{\bv}&  \colhead{$\xi$}&  \colhead{$\sigma_{\xi}$}&
\colhead{$\eta$}&  \colhead{$\sigma_{\eta}$}}
\startdata
\nl
2&ID-31&15.47&1.09&-146.5100&0.0009&34.1564&0.0012\nl
3&ID-32&14.81&0.59&-142.1551&0.0009&-3.6702&0.0013\nl
4&ID-33&14.64&1.01&-170.9684&0.0011&-52.0054&0.0016\nl
5*&ID-30&13.17&1.05&0.0000&0.0008&0.0000&0.0009 \nl
6&ID-34&14.09&1.35&-72.6553&0.0007&94.8155&0.0007\nl
8&ID-36&11.55&0.32&-37.0964&0.0006&90.0358&0.0007\nl
\enddata
\tablenotetext{*} {RA, Dec =  269.4768708, +4.6805700 (J2000)}
\end{deluxetable}
\end{center}

\begin{center}
\begin{deluxetable}{cccccccc}
\tablewidth{0in}
\tablecaption{Proxima Cen Reference Frame  \label{tbl-5}}
\tablehead{\colhead{Star} &  \colhead{ID}&
\colhead{V}&   \colhead{\bv}&  \colhead{$\xi$}&  \colhead{$\sigma_{\xi}$}&  \colhead{$\eta$}&  \colhead{$\sigma_{\eta}$}}
\startdata
\nl
2&ID-13&15.10&0.72&-18.4819&0.0008&-74.9461&0.0014\nl
3&ID-14&15.76&0.87&-3.8168&0.0017&-32.3059&0.0021\nl
4*&ID-15&14.36&0.66&0.0000&0.0005&0.0000&0.0008\nl
5&ID-16&14.58&0.52&70.4847&0.0009&-113.6714&0.0017\nl
6&ID-17&15.32&0.73&15.6880&0.0018&-3.8986&0.0022\nl
7&ID-18&15.14&1.53&-3.6323&0.0007&27.5414&0.0011\nl
9&ID-19&14.30&1.05&131.3827&0.0008&63.0020&0.0013\nl
\tablenotetext{*} {RA, Dec =  217.435381, -62.691332 (J2000)}
\enddata
\end{deluxetable}
\end{center}
\clearpage 
\begin{center}
\begin{deluxetable}{cccccccccccc}
\tablewidth{0in}
\tablecaption{Barnard's Star Reference Frame Absolute Parallax  \label{tbl-6}}
\tablehead{\colhead{Star} &  \colhead{ID}&
\colhead{V}&  \colhead{Sp.T}&  \colhead{\bv}&  \colhead{M$_V$}&  \colhead{(B-V)$_o$}&  \colhead{E$_{B-V}$}&  \colhead{A$_V$}&  \colhead{m-M}&  \colhead{dist(pc)}&  \colhead{$\pi$(")}}
\startdata
2&ID-31&15.47&K0V&1.09&5.9&0.81&0.28&0.88&9.57&547&0.0018\nl
3&ID-32&14.81&F0III&0.59&1.2&0.3&0.29&0.90&13.61&3480&0.0003\nl
4&ID-33&14.64&G9V&1.01&5.7&0.77&0.24&0.74&8.94&437&0.0023\nl
5&ID-30&13.17&G9IV&1.05&3.1&0.85&0.20&0.63&10.07&773&0.0013\nl
6&ID-34&14.09&G9III&1.35&0.75&0.97&0.38&1.17&13.34&2714&0.0004\nl
8&ID-36&11.55&A9III&0.32&1.35&0.26&0.06&0.19&10.2&1005&0.0010\nl
\enddata
\end{deluxetable}
\end{center}
\clearpage
\clearpage

\begin{center}
\begin{deluxetable}{lll}
\tablecaption{Parallax and Proper Motion \label{tbl-7}}
\tablewidth{0in}
\tablehead{\colhead{} &  \colhead{Proxima Cen}&
\colhead{Barnard's Star}}
\startdata
{\it HST} study duration  &5.6 y  & 3.3 y\nl
number of observation sets    & 59  &  34 \nl
ref. stars $ <V> $ &  14.5  &  14.2 \nl
ref. stars $ <B-V> $ &  +0.87  &  +0.85 \nl
Relative Parallax & 767.7 $\pm$ 0.2 & 544.2 $\pm$ 0.2 mas\nl
corr to absolute & 1.0 $\pm$ 0.2 &1.2 $\pm$ 0.2   mas\nl
{\it HST} Absolute Parallax  & 768.7 $\pm$ 0.3 & 545.4 $\pm$ 0.3  mas\nl
Yale Parallax Catalog (1995)& 769.8 $\pm$ 6.1 & 545.6 $\pm$ 1.3   mas\nl
{\it HIP}  Parallax  & 772.3 $\pm$ 2.4 & 549.3 $\pm$ 1.6  mas\nl
{\it HST} Proper Motion &3851.7 $\pm$ 0.1 &10370.0 $\pm$ 0.3 mas y$^{-1}$ \nl
 \indent in p.a. & 78\fdg46 $\pm$ 0\fdg03 &4\fdg45 $\pm$ 0\fdg1 \nl
{\it HST} Secular Acceleration & - & 12 $\pm$ 4 mas y$^{-2}$ \nl
{\it HIP}  Proper Motion &3852.9 $\pm$ 2.3 &10368.6 $\pm$ 2.1 mas y$^{-1}$ \nl
\indent  in p.a. &78\fdg50 $\pm$0\fdg03 &4\fdg42 $\pm$ 0\fdg07 \nl

\enddata
\end{deluxetable}
\end{center}

\clearpage

%
%

\clearpage

\begin{figure}
\epsscale{.6}
\plotone{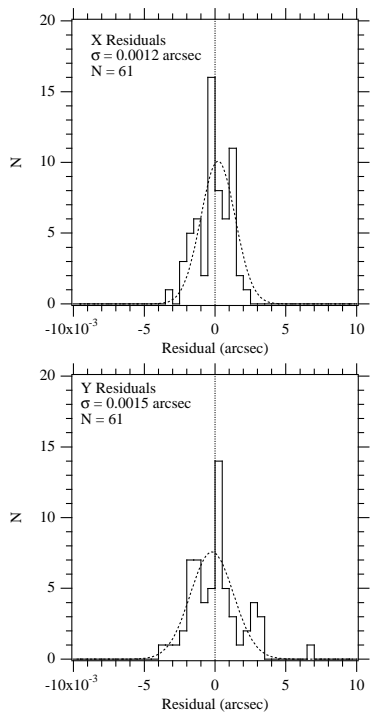}
\caption{Histograms of Lateral Color calibration observation residuals to the model (equations 5 and 6). } \label{fig-1}
\end{figure}
\clearpage
\begin{figure}
\epsscale{1.0}
\plotone{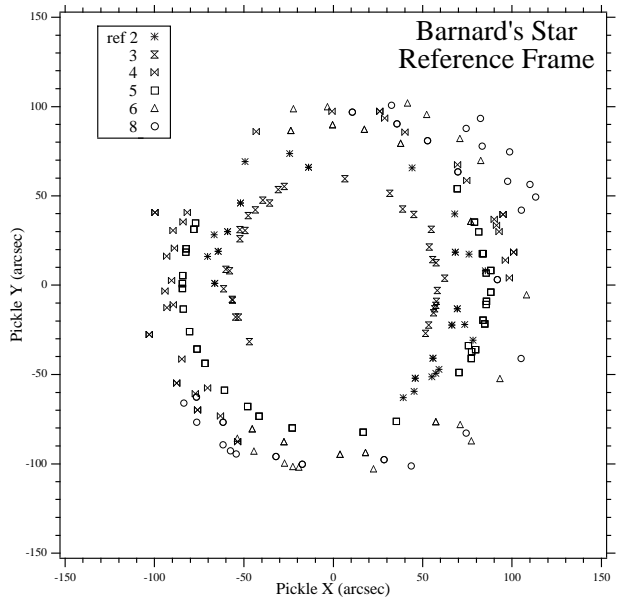}
\caption{Barnard's Star reference star observations in FGS 3 pickle coordinates. The symbol
shape identifies each star (listed in Table 4). } \label{fig-2}
\end{figure}
\clearpage
\begin{figure}
\plotone{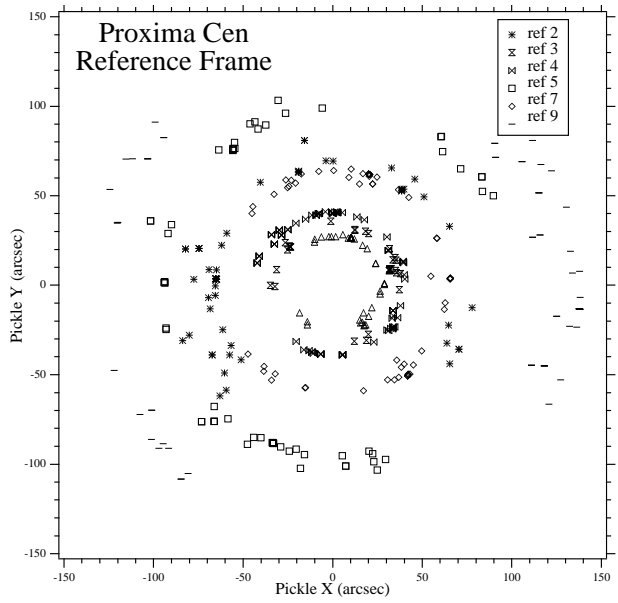}
\caption{Proxima Cen reference star observations in FGS 3 pickle coordinates. The symbol
shape identifies each star (listed in Table 5). } \label{fig-3}
\end{figure}
\clearpage
\begin{figure}
\epsscale{0.8}
\plotone{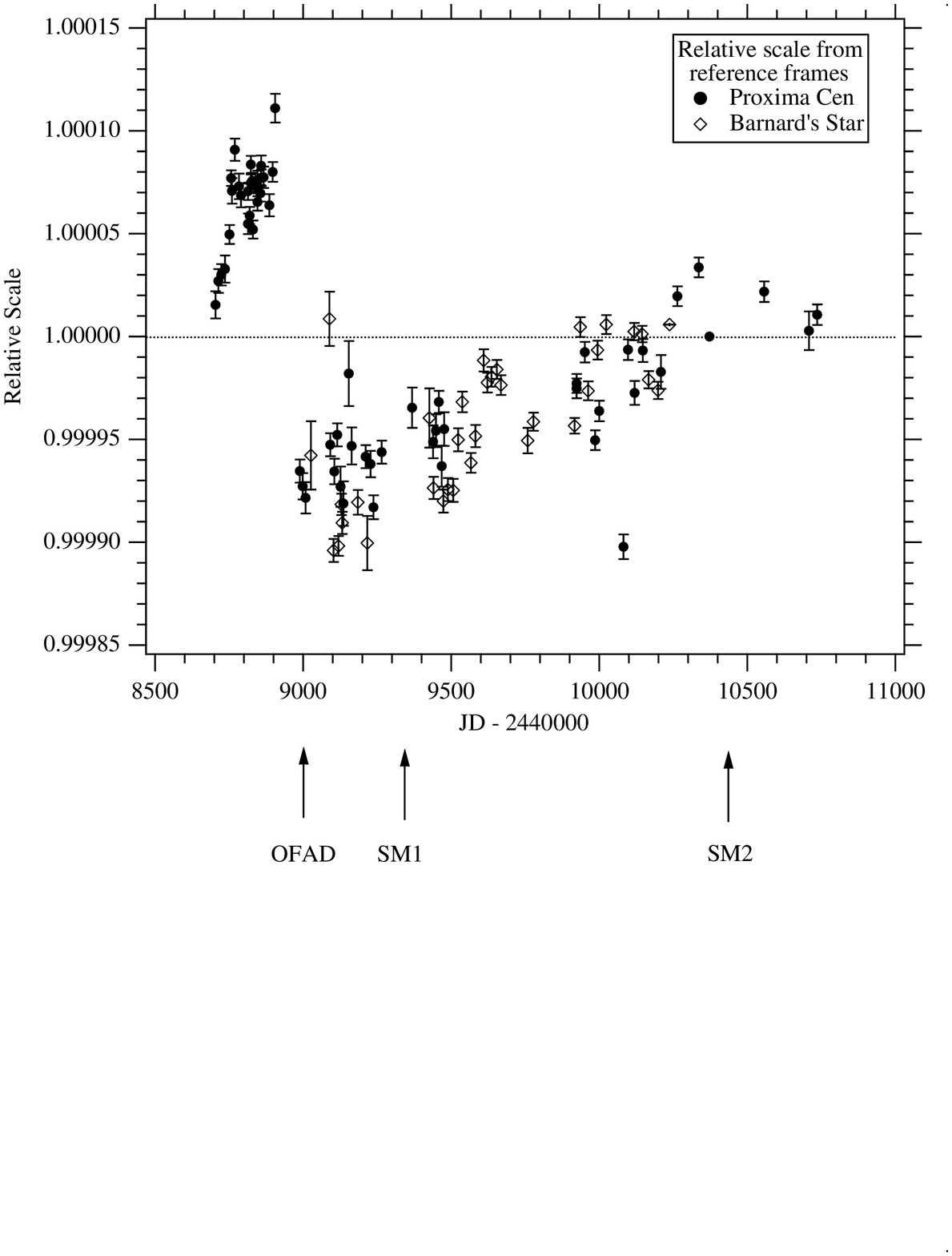}
\caption{Comparison of scales derived from equation 9 for Prox
Cen and Barnard's Star reference frames. Major events are labeled: OFAD, first full-up
astrometric calibration of FGS 3; SM1 and SM2, servicing missions.} \label{fig-4}
\end{figure}
\clearpage
\begin{figure}
\plotone{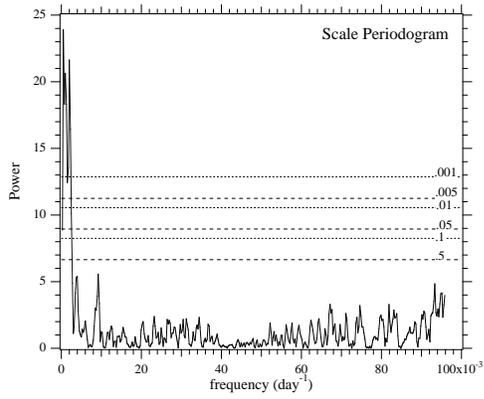}
\caption{Power spectrum of scales derived from equation 9 for Prox
Cen and Barnard's Star reference frames. The horizontal lines are levels of false-positive probability (equation A-4). } \label{fig-5}
\end{figure}
\clearpage

\begin{figure}
\epsscale{.6}
\plotone{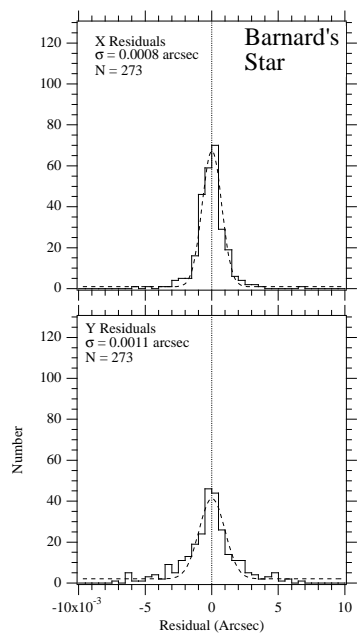}
\caption{Histograms of X and Y residuals obtained from modeling the Barnard's Star reference frame with equations 7 and 11. Distributions are fit with gaussians characterized by $\sigma$ as indicated.} \label{fig-6}
\end{figure}
\clearpage
\begin{figure}
\plotone{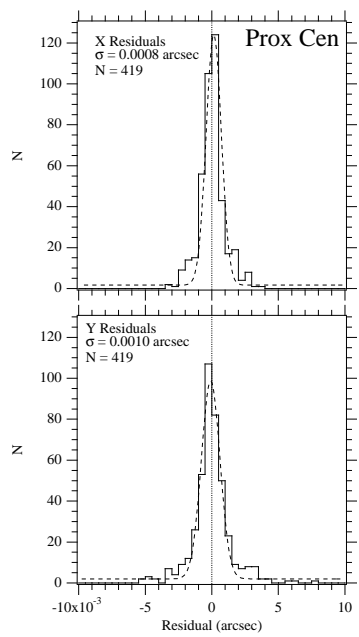}
\caption{Histograms of X and Y residuals obtained from modeling the Proxima Cen reference frame with equations 7 and 8. Distributions are fit with gaussians characterized by $\sigma$ as indicated.} \label{fig-7}
\end{figure}
\clearpage
\begin{figure}
\epsscale{1.0}
\plotone{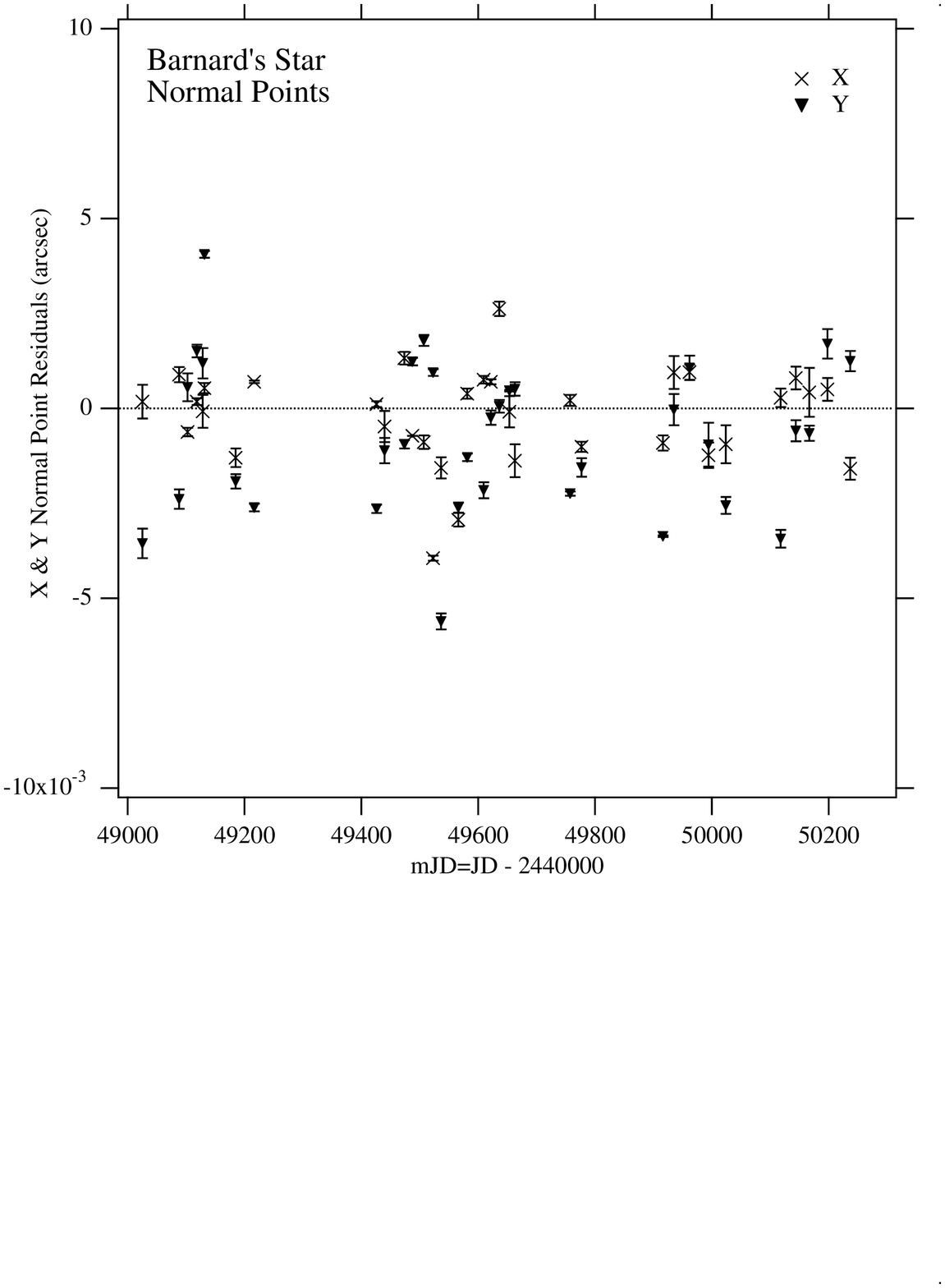}
\caption{ Residual normal points in X and Y for Barnard's Star plotted against time. } \label{fig-8}
\end{figure}
\clearpage
\begin{figure}
\plotone{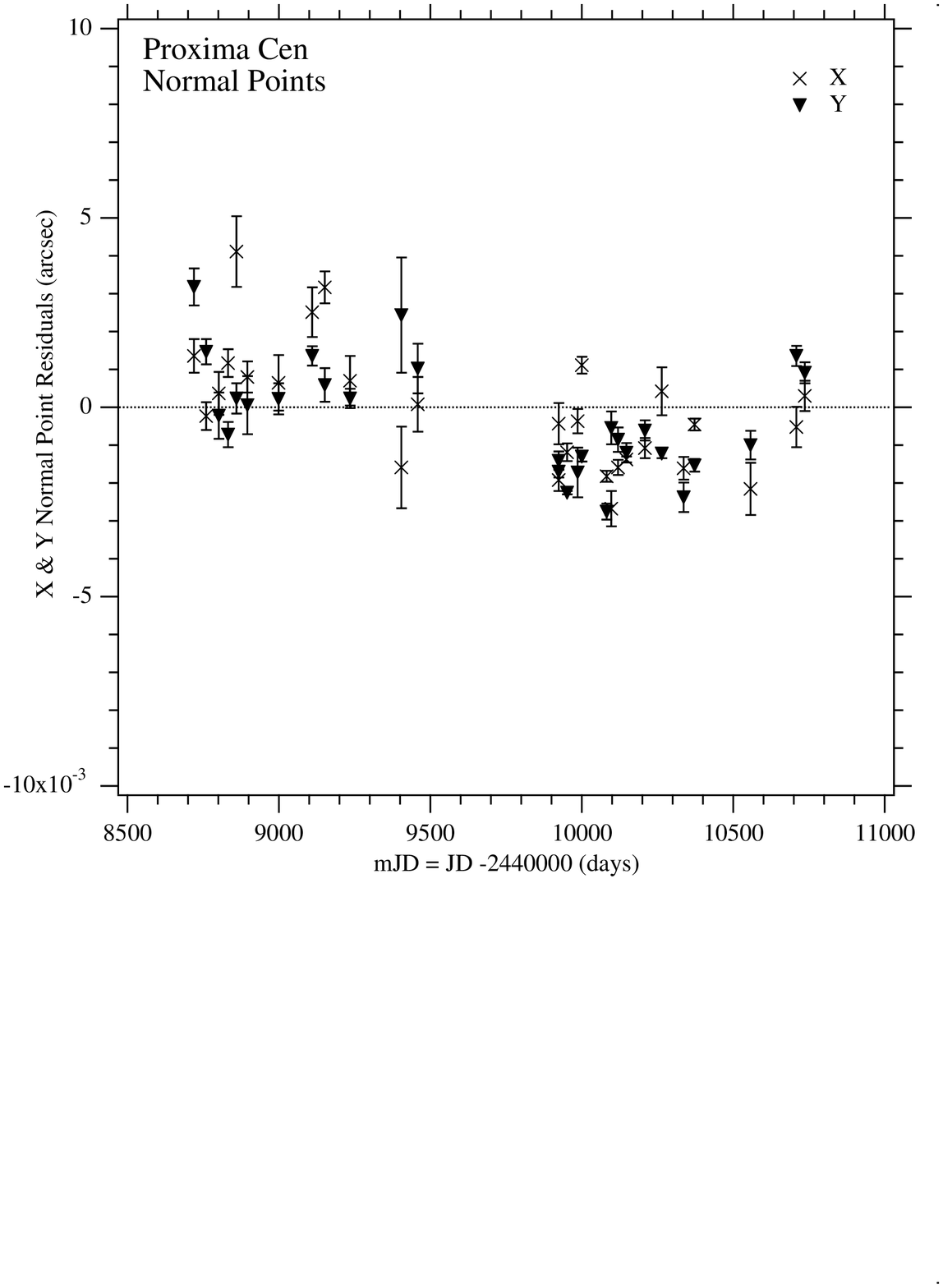}
\caption{ Residual normal points in X and Y for Proxima Cen plotted against time. } \label{fig-9}
\end{figure}
\clearpage

\begin{figure}
\epsscale{1.0}
\plotone{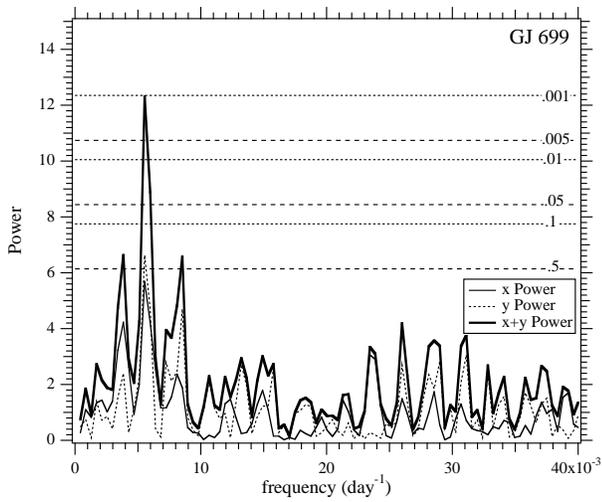}
\caption{Lomb-Scargle periodogram of the Figure~\ref{fig-8} normal points for Barnard's Star. The solid line is the sum of the X and Y power spectra. Horizontal lines denote levels of false positive probability from equation A-4. There is a significant signal at $f\sim 182 d^{-1}$.} \label{fig-10}
\end{figure}
\clearpage
\begin{figure}
\plotone{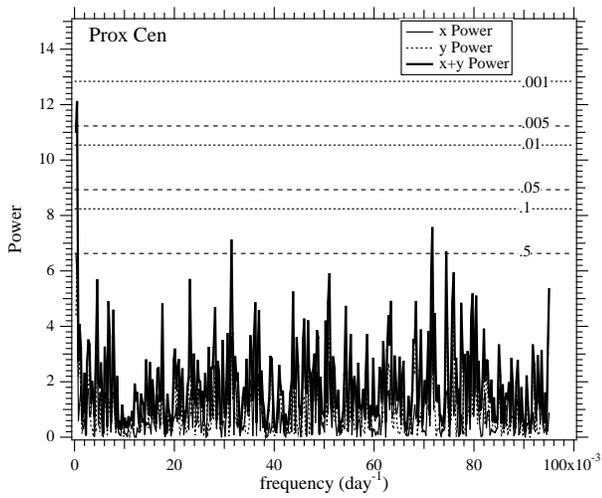}
\caption{ Lomb-Scargle periodogram of the Proxima Cen normal points in Figure~\ref{fig-9}. The solid line is the sum of the X and Y power spectra. Horizontal lines denote levels of false positive probability. Note the significant signal at $f\sim 1800 d^{-1}$.} \label{fig-11}
\end{figure}
\clearpage
\begin{figure}
\epsscale{1.0}
\plotone{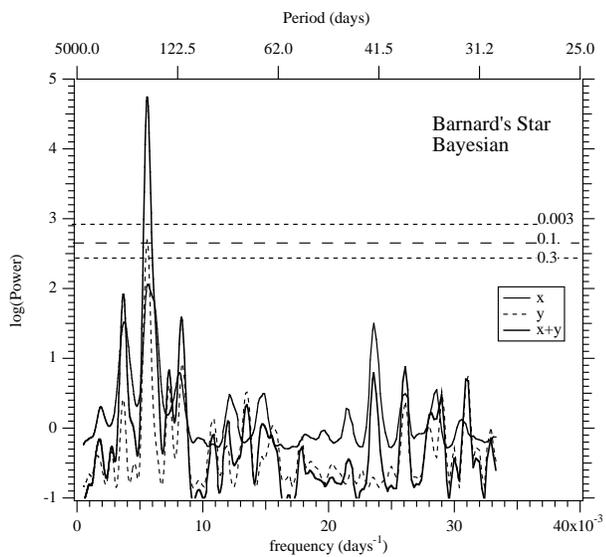}
\caption{Bayesian periodogram of the Figure~\ref{fig-9} normal points for Barnard's Star. The solid line is the sum of the X and Y power spectra. Horizontal lines denote levels of false-positive probability from Monte-Carlo simulations described in \S A1.2 of the Appendix. This approach also detects significant signal at $f\sim 182 d^{-1}$.} \label{fig-12}
\end{figure}
\clearpage
\begin{figure}
\epsscale{0.6}
\plotone{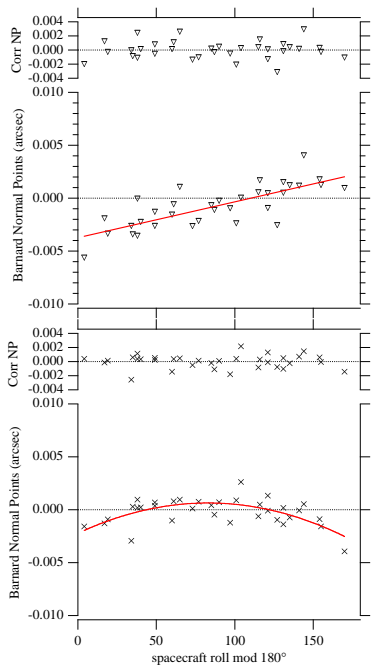}
\caption{ Barnard's Star normal points plotted against {\it HST} roll, modulo 180\arcdeg. There are obvious trends, linear in Y, quadratic in X. The residuals
(CorrNP) are adopted as the corrected normal points.} \label{fig-13}
\end{figure}
\clearpage
\begin{figure}
\epsscale{1.0}
\plotone{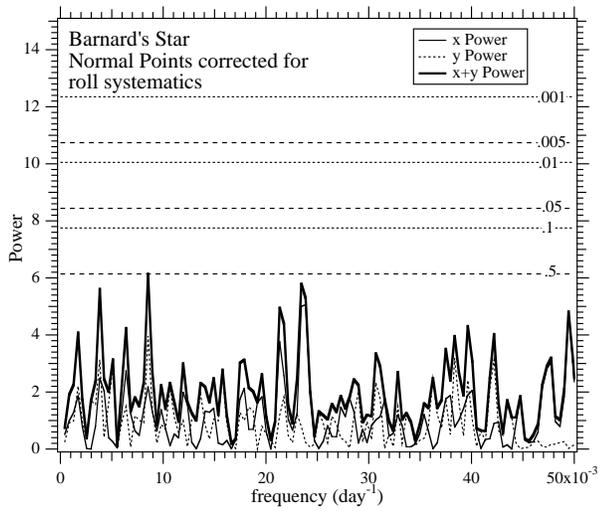}
\caption{ Lomb-Scargle periodogram of Barnard's Star (CorrNP) normal points corrected for systematic error depending on {\it HST} roll. The solid line is the sum of the X and Y power spectra. Horizontal lines denote levels of false positive probability. The correction has effectively removed the power at $f\sim 182 d^{-1}$ seen in Figure~\ref{fig-10}.} \label{fig-14}
\end{figure}
\clearpage
\begin{figure}
\epsscale{1.0}
\plotone{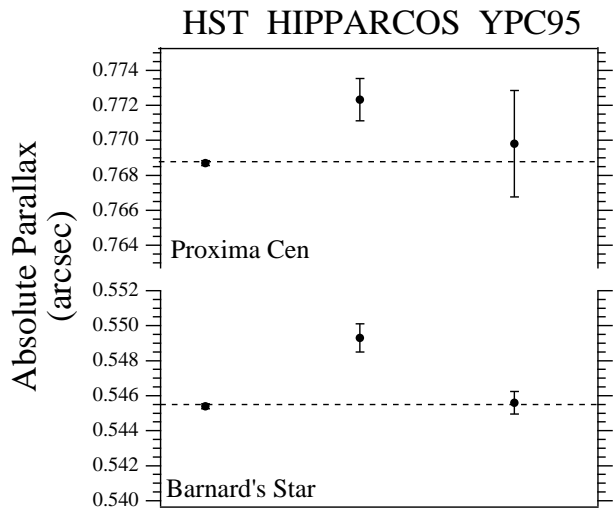}
\caption{Absolute parallaxes for Proxima Cen and Barnard's Star determined by
{\it HST} (left) and {\it HIPPARCOS} (center), compared with the YPC95 (right). The horizontal dashed lines are the weighted averages of the three independent determinations. } \label{fig-15}
\end{figure}
\clearpage
\begin{figure}
\epsscale{1.0}
\plotone{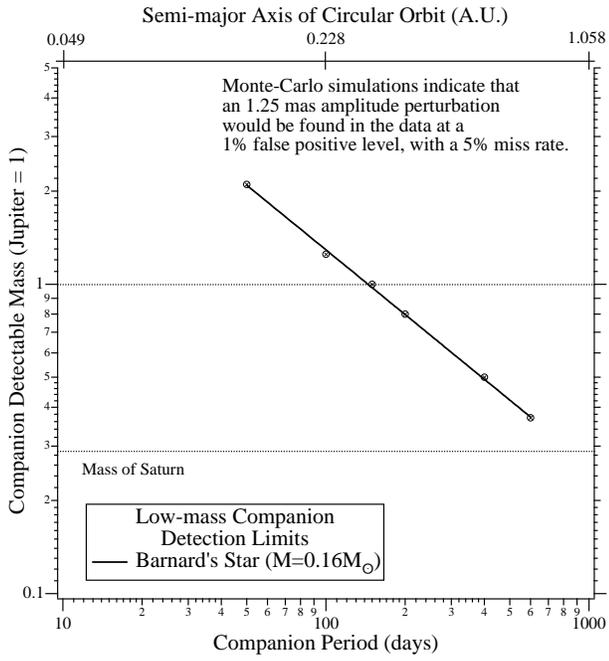}
\caption{Barnard's Star companion detection mass limits. The top axis gives the size of the semimajor axis of a 1 $M_{Jup}$ secondary. Monte-Carlo simulations indicate that
an 1.25 mas amplitude perturbation 
would be found in the data at a 
1\% false positive level, with a 5\% miss rate. } \label{fig-16}
\end{figure}
\clearpage
\begin{figure}
\epsscale{1.0}
\plotone{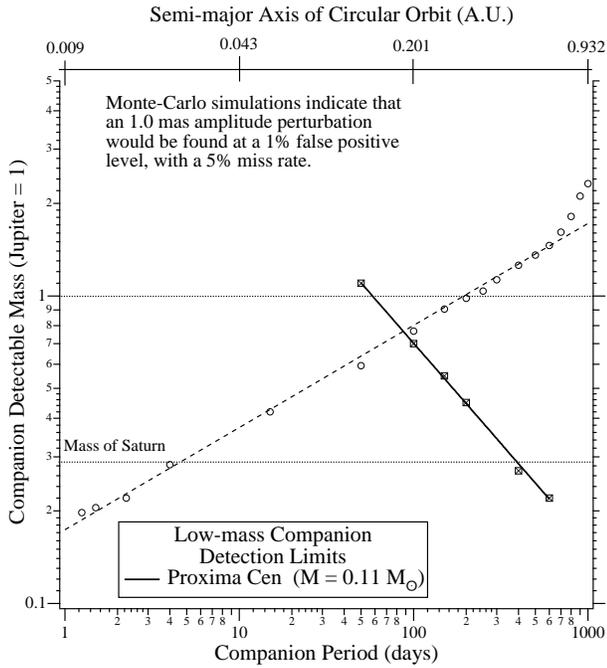}
\caption{ Proxima Cen companion detection mass limits. The top axis gives the size of the semimajor axis of a 1 $M_{Jup}$ secondary. Monte-Carlo simulations indicate that
an 1.0 mas amplitude perturbation 
would be found at a 1\% false positive 
level, with a 5\% miss rate. The dashed line presents the limits determined by the radial velocity program of ~\cite{Kur99}.} \label{fig-17}
\end{figure}



\end{document}